\documentstyle[12pt]{article}
\begin{document}
\input{epsf}
\Large
\bf
\begin{center}   
      {Application of the  Two-Scale Model}
\end{center}
\begin{center}
      {to the HERMES Data on Nuclear Attenuation}
\end{center}
\begin{center}
      
\end{center}

\normalsize
\begin{center}
      N.~Akopov,  L.~Grigoryan\footnote{*) supported by DESY, Deutsches Elektronen Synchrotron}, Z.~Akopov
\end{center}

\begin{center}
      Yerevan Physics Institute, Br.Alikhanian 2, 375036 Yerevan, Armenia
\end{center}

\begin{center}
\end{center}

\begin {abstract}


\hspace*{1em}The Two-Scale Model and its improved version were used to 
perform the
 fit to the HERMES data for $\nu$ (the virtual photon energy) and z (the fraction of
 $\nu$ carried by hadron) dependencies of nuclear 
 multiplicity ratios for $\pi^+$ and $\pi^-$ mesons electro-produced on two nuclear 
 targets ($^{14}$N
 and $^{84}$Kr). 
The quantitative criterium $\chi ^2$ was used for the first time to analyse
the results of the model fit to the nuclear multiplicity ratios data.
The two-parameter's fit gives satisfactory agreement
 with the HERMES data. Best values of the parameters were then used to
 calculate the $\nu$- and $z$ - dependencies of nuclear attenuation for $\pi^0$, K$^+$, K$^-$ and $\bar{p}$ produced
 on $^{84}$Kr target, 
 and also make a predictions for $\nu$, z and the Q$^2$ (the photon virtuality) - dependencies of nuclear attenuation data for 
 those identified hadrons and nuclea, that will be published by HERMES.
\end {abstract}
\bf   
\section{Introduction}
\normalsize 
\hspace*{1em} Studies of hadron production in deep inelastic 
semi-inclusive
 lepton-nucleus scattering (SIDIS) offer a possibility to investigate the
 quark (string, color dipole) propagation in dense nuclear matter and
 the space-time evolution of the hadronization process. It is well-known
 from QCD, that confinement forbids existence of an isolated color
 charge (quark, antiquark, etc.). Consequently, it is clear that after
 Deep Inelastic Scattering (DIS) of lepton on intra-nuclear nucleon, the
 complicated colorless pre-hadronic system arises. 
 Its propagation in the
 nuclear environment involves processes like multiple interactions with
 the surrounding medium and induced gluon radiation. If the final hadron
 is formed inside the nucleus, the hadron can interact via the relevant
 hadronic cross section, causing further reduction 
 of the hadron yield~\cite{A1}. QCD at present can not describe the process of
 quark hadronization because of the major role of "soft" interactions.
 Therefore, the investigation of quark hadronization is of basic importance 
 for development of QCD.
  For this purpose we investigate in this paper the Nuclear Attenuation (NA), which is the ratio of the 
differential multiplicity on nucleus to
 that on deuterium.
  At present there exist numerous phenomenological models for investigation of
 the NA problem~\cite{A2}-\cite{A14}.
  In this work we use the Two-Scale Model~\cite{A4} and its improved version
 to perform the fit to the HERMES NA data ~\cite{A15,A15a}. For the fitting purposes we use the more precise
 part of data, including data for $\nu$- and z - dependencies of NA of
 $\pi^+$ and $\pi^-$ mesons on two nuclear targets ($^{14}$N and 
 $^{84}$Kr).
The $\nu$- and z - dependencies of NA for $\pi^0$, K$^+$, K$^-$ and 
antiproton, produced on  $^{84}$Kr target  we describe
 with best values of parameters obtained from the above mentioned fit. The best set of
 parameters are used also for prediction of $\nu$, z and Q$^2$-
 dependencies of NA for the data on those identified hadrons and targets that will be published
 soon by HERMES~\cite{dis03}.
  The remainder of the paper is organized as follows. In section 2 we
 briefly remind about the Two-Scale Model. In section 3 we discuss the possibility 
 of inclusion of the Q$^2$-dependence in the Two-Scale Model . In section 4
 we describe the scheme we used to improve the Two-Scale Model, substituting the step-by-step increase 
 of the string-nucleon cross section by a smooth raising function. In section 5 we present the 
 results of the model fit to the HERMES data. Our conclusions are given in section 6.

\bf
\section{The Two-Scale Model}
\normalsize
\hspace*{1em}The Two-Scale Model is a string model, which was proposed 
by EMC~\cite{A4} 
 and used for the description of their experimental data.
  Basic formula is:
  
\begin{eqnarray}
R_A = 2\pi\int_0^{\infty} bdb \int_{-\infty}^{\infty}dx \rho(b,x) [1-\int_x^{\infty} dx' \sigma^{str}(\Delta x)\rho (b,x')]^{A-1}  
\end{eqnarray}

where b - impact parameter,
 x - longitudinal coordinate of the DIS point,
 x$'$- longitudinal coordinate of the string-nucleon interaction point,
$\sigma^{str}$($\Delta$x) - the string-nucleon cross section on distance
 $\Delta$x = x'-x from DIS point,
 $\rho(b,x)$ - nuclear density function, A - atomic mass number.
 
\begin{figure}
\begin{center}
\epsfxsize=8cm
\epsfysize=7cm
\epsfbox{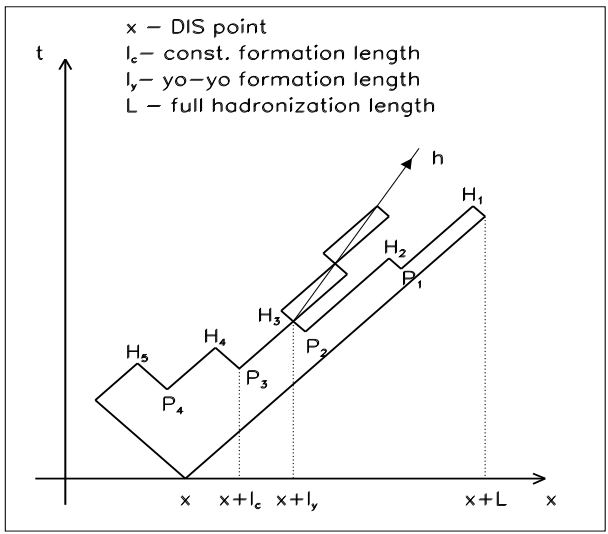}
\end{center}
\caption{\label{xx1}  
Space-time structure of hadronization in the string model. The two
constituents of the hadron are produced at different points. The constituents of
the hadron h are created at the points P$_2$ and P$_3$. They meet at H$_3$ to form the
hadron.}
\end{figure}
  The model contains two scale (see Fig. 1):
 $\tau_c$ ($l_c$) - constituent formation time (length), and
 $\tau_h$ ($l_h$) - yo-yo formation time (length),
\footnote{in relativistic units ($\hbar$ = c = 1,
where $\hbar = h/2\pi$ is the Plank reduced constant and
c - speed of light) $\tau_i = l_i$, i=c,h because
partons and hadrons move with near light speeds.}
 yo-yo formation means, that
the  colorless system with valence content and quantum numbers of final 
 hadron  arises, but without its "sea" partons.
  The simple connection exists between $\tau_h$ and $\tau_c$

\begin{eqnarray}
           \tau_h - \tau_c = z\nu/\kappa ,                     
\end{eqnarray}

 where
 z$ = E_h/\nu$, E$_h$ and $\nu$ are energies of final hadron and virtual
 photon correspon-\\dingly, $\kappa$ - string tension (string constant).
Further we will use two different expressions for $\tau_c$.
 The expression for $\tau_c$ obtained for hadrons containing leading
 quark~\cite{A16}:

\begin{eqnarray}
     \tau_c =(1 - z)\nu/\kappa .                               
\end{eqnarray}

 The expression for average value of $\tau_c$, which was obtained in~\cite{A5,A17} in
 framework of the standard Lund model~\cite{A18}:

\begin{eqnarray}
    \tau_c = \int_0^{\infty} ldlD_c(L,z,l)/\int_0^{\infty} dlD_c(L,z,l),                                                   
\end{eqnarray}

 where D$_c$(L, z, l) is the distribution of the constituent formation 
 length l of hadrons carrying momentum z. This distribution
is:

\begin{eqnarray}
    D_c(L,z,l) =
    L(1+C)\frac{l^C}{(l+zL)^{C+1}}(\delta(l-L+zL)+\frac{1+C}{l+zL})\theta(l)\theta(L-zL-l) ,
\end{eqnarray}

 where L = $\nu/\kappa$, and parameter C=0.3.
  The path traveling by string between DIS and interaction points
 is $\Delta$x = x'-x. The string-nucleon cross section is:
\vskip 0.5cm 
\begin{eqnarray}
\sigma^{str}(\Delta x) = \theta(\tau_c - \Delta x)\sigma_q + \theta(\tau_h - \Delta x)\theta(\Delta x - \tau_c)\sigma_s + 
\theta(\Delta x - \tau_h)\sigma_h
\end{eqnarray}	
\vskip 0.5cm 		  
where $\sigma_q$, $\sigma_s$ and $\sigma_h$ are the cross sections for
 interaction with nucleon of initial string, open string (which becomes 
 one of the hadron quarks being looked at) and final hadron respectively
 (see Fig.2 a)).
\section{Inclusion of the Q$^2$-dependence\\
 in Two-Scale Model.}

  The Two-Scale Model~\cite{A4} does not contain direct Q$^2$-dependence and
 operates with the average values of cross sections:

\begin{eqnarray}
    \sigma_q = \sigma_q(\hat{Q}^2); \hskip 1cm \sigma_s = \sigma_s(\hat{Q}^2_{\tau_c}),
\end{eqnarray}

 where $\hat{Q}^2$ is average value of Q$^2$ obtained in experiment for
 initial state and $\hat{Q}^2_{\tau_c}$ is the same for open string, or
for time $\tau_c$ after DIS point. Obviously that $\hat{Q}^2_{\tau_c}$ is
 smaller than $\hat{Q}^2$, because after DIS the string radiates gluons
 and diminishes its virtuality.
  QCD predicts the Q$^2$-dependence of string-nucleon cross section in the
 form~\cite{A19,A20}:

\begin{eqnarray}
   \sigma_q(Q^2)\sim 1/Q^2;\hskip 1cm \sigma_s(Q^2_{\tau_c})\sim 1/Q^2_{\tau_c}.
\end{eqnarray}

 Using this prediction we can write the cross section for initial
 string as

\begin{eqnarray}
   \sigma_q(Q^2) = (\hat{Q}^2/Q^2)\sigma_q(\hat{Q}^2).                
\end{eqnarray}
In the same way can be written the expression for open string
 cross section

\begin{eqnarray}
 \sigma_s(Q^2_{\tau_c}) = (\hat{Q}^2_{\tau_c}/Q^2_{\tau_c})\sigma_s(\hat{Q}^2_{\tau_c}),
\end{eqnarray}

 where Q$^2_{\tau_c}$ is the virtuality of string for time $\tau_c$
 after DIS point, and $\hat{Q}^2_{\tau_c}$ is the same for average value 
 of Q$^2$. For estimation of ratio $\hat{Q}^2_{\tau_c}/Q^2_{\tau_c}$
 we adopt the scheme given in Ref.~\cite{A21,A22}. In according with
 this scheme, for the time t the quark decreases its virtuality
 from the initial one, Q$^2$, to the value Q$^2$(t)

\begin{eqnarray}
    Q^2(t) = \nu(t)\frac{Q^2}{\nu(t)+tQ^2},
\end{eqnarray}

where $\nu$(t) = $\nu$ -$\kappa t$.
 The calculations shown, that for HERMES kinematics \\
 (1.2$<$Q$^2<$9.5~GeV$^2$ and $\hat{Q}^2$=2.5~GeV$^2$), values for ratio $\hat{Q}^2_{\tau_c}/Q^2_{\tau_c}$ are
 close to 1 (for $\tau_c$ in form of (3) it changes in region 0.97$\div$1.04
 and  in form of (4) in region 0.92$\div$1.12). This means that $\sigma_s$
 is practically constant.

\section{Improved version of Two-Scale Model}

  In the Two-Scale Model the string-nucleon cross section is a function 
  which jumps in points $\Delta$x = $\tau_c$ and $\tau_h$. In reality
 the cross section increases smoothly until it reaches the size of hadronic 
 cross section. That is why we need to improve the model in order
 to obtain the smooth increase of the cross section (see Fig. 2).
\begin{figure}
\begin{center}
\epsfxsize=10cm
\epsfysize=8cm
\epsfbox{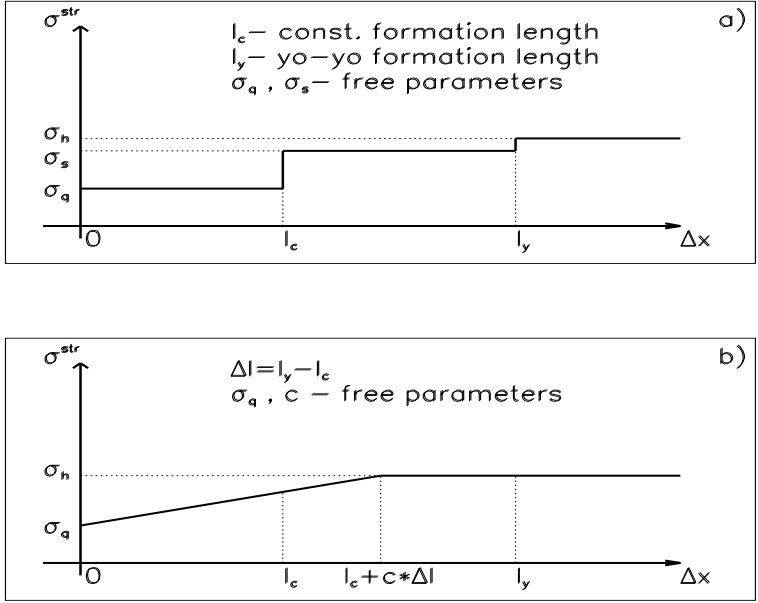}
\end{center}
\caption{\label{xx2} a) The behaviour of the string-nucleon cross section as 
a function of distance in the Two-Scale Model. 
b) The same as
in a) for improved Two-Scale Model with taking into account more realistic
smoothly increasing string-nucleon cross section.
}
\end{figure}
  We introduce the parameter $c$ (0$<c<$1)
 in order to take into account the well known fact, that string
 starts to interact with hadronic cross section soon after creation 
 of the first constituent quark of the final hadron, before
 creation of second constituent. The string-nucleon cross section
 starts to increase from DIS point, and reaches the value of the
 hadron-nucleon cross section at $\Delta x$ = $\tau$. However, in that
 case one cannot deduce the exact form of $\sigma^{str}$ from perturbative 
 QCD, at least in region $\Delta$x $\sim$ $\tau$. This means, that
 some model for the shrinkage-expansion mechanism has to be invented. 
 We use four versions for $\sigma^{str}$. Two versions we took
 from Ref.~\cite{A23}. The first version is based on quantum diffusion:
 
\begin{eqnarray} 
\sigma^{str}(\Delta x) = \theta (\tau- \Delta x)[\sigma_q + (\sigma_h-\sigma_q)
\Delta x/ \tau ] + \theta (\Delta x - \tau)\sigma_h
\end{eqnarray}

where $\tau$ = $\tau_c$ + c$ \Delta \tau$, $\Delta \tau$ = $\tau_h$ - $\tau_c$,
 
The second version follows from naive parton case:

\begin{eqnarray}
 \sigma^{str}(\Delta x) = \theta (\tau- \Delta x)[\sigma_q +
 (\sigma_h-\sigma_q)(\Delta x/\tau)^2 ] + \theta (\Delta x - \tau)\sigma_h
\end{eqnarray}

 We used also two other expressions for $\sigma^{str}$~\cite{A2,A6}:

\begin{eqnarray} 
 \sigma^{str}(\Delta x) = \sigma_h - (\sigma_h-\sigma_q)exp(-\Delta x/\tau) 
\end{eqnarray}

and:

\begin{eqnarray}
 \sigma^{str}(\Delta x) = \sigma_h - (\sigma_h-\sigma_q)exp(-(\Delta x/\tau)^2) 
\end{eqnarray}

 One can easily note that at $\Delta x/\tau\ll1$ the expressions (14) and (15)
 turn into (12) and (13), correspondingly.

\section{Results}

 In this work we have formulated one of possible improvements of the Two-Scale model and
performed a fit to the HERMES data~\cite{A15,A15a}.
 Only the data for NA for $\nu$- and z - dependencies 
 of $\pi^+$ and $\pi^-$ mesons on $^{14}$N and $^{84}$Kr nuclea were used for the actual fit.
 Furthermore, NA for $\nu$- and z - dependencies of other hadrons produced on $^{84}$Kr target were
 calculated. Also based on the best fit parameters one can make different predictions for $\nu$, z  and 
 Q$^2$- dependencies 
 for those identified hadrons and nuclea, that will be published by HERMES~\cite{dis03}.
The string tension (string constant) was fixed at a static value
 determined by the Regge trajectory slope~\cite{A22,A24}

\begin{eqnarray}
     \kappa = 1/(2\pi\alpha'_R) = 1 GeV/fm                    
\end{eqnarray}

 We use the following Nuclear Density Functions (NDF):
 
 For $^4$He and $^{14}$N we use the Shell Model~\cite{A25}, according to which four nucleons ( 
 two protons and two neutrons), fill the s - shell,
and other A-4 nucleons are on the p - shell:

\begin{eqnarray}
\rho(r)=\rho_0(\frac{4}{A} +\frac{2}{3}\frac{(A-4)}{A}\frac{r^2}{r_A^2})exp(-\frac{r^2}{r_A^2}),
\end{eqnarray}
  where $r_A$=1.31~fm for $^4$He and $r_A$=1.67~fm for $^{14}$N.
  
 For $^{20}$Ne, $^{84}$Kr and $^{131}$Xe we use Woods-Saxon distribution
\begin{eqnarray}
\rho(r) = \rho_0 /(1+exp((r-r_A)/a)).                        
\end{eqnarray}
 The three sets of NDF were used for the fitting with the following corresponding
 parameters:
 
 First set (NDF=1)~\cite{A26}. 
\begin{eqnarray}
 a=0.54~fm;\hskip 1cm r_A =(0.978 + 0.0206A^{1/3})A^{1/3}~fm            
\end{eqnarray}
 Second set (NDF=2)~\cite{A27}
\begin{eqnarray}
a=0.54~fm;\hskip 1cm r_A =(1.19A^{1/3} -1.61/A^{1/3})~fm 
 \end{eqnarray}
 Third set (NDF=3)~\cite{A28}
\begin{eqnarray}
 a=0.545~fm;\hskip 1cm r_A =1.14A^{1/3}~fm.                             
 \end{eqnarray}
where $\rho_0$ are determined from normalization condition:
\begin{eqnarray}
           \int{d^3r \rho(r)} = 1                         
\end{eqnarray}

 Parameter $a$ is practically the same for all three sets, radius
 $r_A$ for the third set is larger approximately on 6\%  than for
 second and first sets.
  From the fit we determined two parameters for Two-Scale Model
                  $\sigma_q$ and $\sigma_s$.
 In case of the improved Two-Scale Model the fitting parameters
 are $\sigma_q$ and c.

 Determination of the parameter $c$ is represented in Section 4. For fitting 
 we used two expressions for $\tau_c$, which are equations (3)
 and (4), and five expressions for $\sigma^{str}(\Delta x)$ (6), (12)-
 (15). The results of the performed fit are presented in Tables 1, 2a and 2b. 
 As we have mentioned above only part of HERMES
experimental data was used for the fitting procedure, including $\nu$ - and z - dependencies of NA
for $\pi^+$ and $\pi^-$ on $^{14}$N and $^{84}$Kr nuclea. For each 
measured bin the information on the values of $\hat{z}$ (averaged over the given
$\nu$ bin) 
in case of $\nu$ dependence, and $\hat{\nu}$ in case of z dependence was
taken from experimental data. Use of this information 
allows to avoid the problem of additional integration over z and $\nu$ in
formulae (1). 

In Table 1 
the best values for fitted parameters, their errors and $\chi^2$/d.o.f. (N$_{exp}$=58,
N$_{par}$=2. N$_{exp}$ and N$_{par}$ are the numbers of experimental points and fitting
parameters which were used.) for the Two-Scale Model are represented. 
Two different expressions for $\tau_c$, and three 
different sets of parameters for NDF ($^{84}$Kr) were used.
Tables 2a and 2b contain the best values for fitted parameters, their errors and
$\chi^2$/d.o.f. (N$_{exp}$=58, N$_{par}$=2) for the Improved Two-Scale
Model. Four different 
expressions for $\sigma^{str}$ were used. Only difference between Tables 2a and 2b
is the form of $\tau_c$. The results
 for Two-Scale Model (Table 1) are qualitatively close to the results 
 of Ref.~\cite{A4}. The values of $\sigma_q \ll \sigma_h$ and $\sigma_s$ are approximately 
 equal to $\sigma_h$. $\sigma_q$ in our case is larger than the same
 in Ref.~\cite{A4}, because $\hat{Q}^2$ for HERMES kinematics is smaller than
 in EMC kinematics.
  The minimum values for $\chi^2$/d.o.f. (best fit) were obtained for
 the Improved Two-Scale Model with the constituent formation time
 $\tau_c$ in form of (3) (see Table 2a).

The results for NA, calculated with the best values of fitting
parameters for improved Two-Scale Model, for $\nu$ and z dependencies
of produced charged pions on $^{14}$N and $^{84}$Kr targets are presented
on Fig. 3.
 
 \begin{figure}
\begin{center}
\epsfxsize=10cm
\epsfysize=12cm
\epsfbox{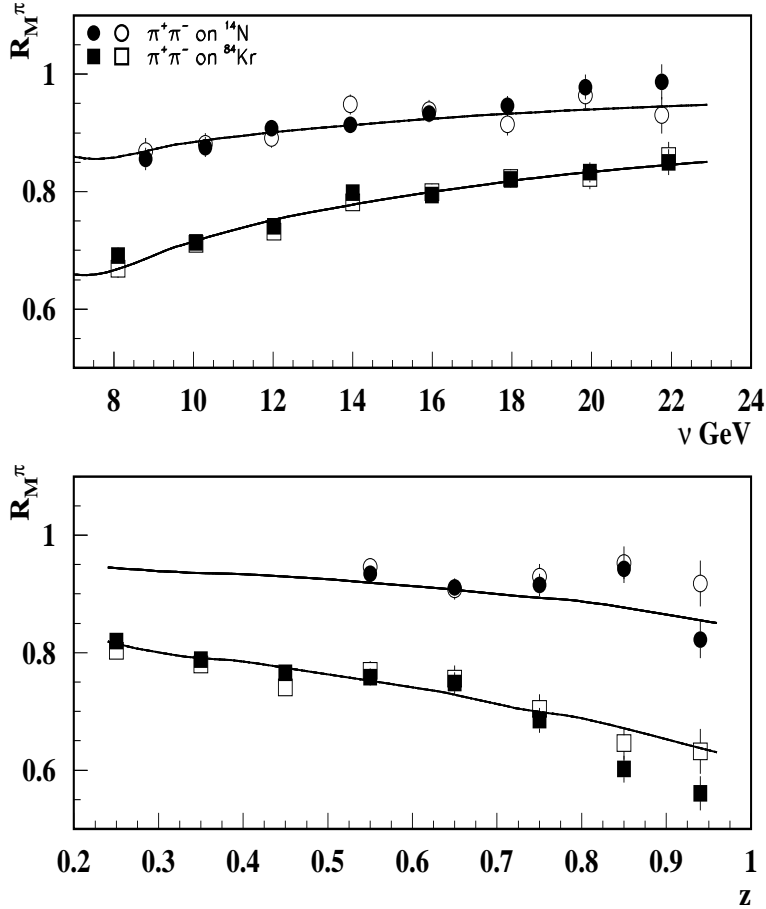}
\end{center}
\caption{\label{xx3}Hadron multiplicity ratio R of charged pions
 for $^{14}$N and $^{84}$Kr nuclea as a function of $\nu$ (upper panel),
z (lower panel). The
theoretical curves correspond the calculations in Improved
Two-Scale Model performed with NDF (17) for $^{14}$N and NDF (19) for $^{84}$Kr and $\sigma^{str}$ (12)
with $\tau_c$ in form (3) for the values of parameters: $\sigma_q$=0.46mb, c=0.32. 
} 
\end{figure}
In Fig. 4 one can see the $\nu$ and z 
 dependencies for all identified hadrons
 produced on $^{84}$Kr target. The values of $\sigma_h$ (hadron-nucleon inelastic cross section) used in this
 work are equal to: $\sigma_{\pi^+} = \sigma_{\pi^-} = \sigma_{\pi^0} = \sigma_{K^-}$ = 20~mb, $\sigma_{K^+}$ = 14~mb and $\sigma_{\bar{p}}$ = 42~mb. 
 The curves correspond to the improved Two-Scale model with the 
 best set of parameters. 
 
In Fig. 5 we present the results of the Two-Scale Model and its improved version in comparison with the experimental data for NA of
charged hadrons on $^{63}$Cu target~\cite{A4} performed in region of $\nu$ and Q$^2$ values higher, than in
HERMES kinematics. In order to compare with the EMC data we redefined $\sigma_q$ to the $\hat{Q}^2_{EMC}$,
according to the expression (9). 

\begin{figure}
\begin{center}
\epsfxsize=10cm
\epsfysize=12cm
\epsfbox{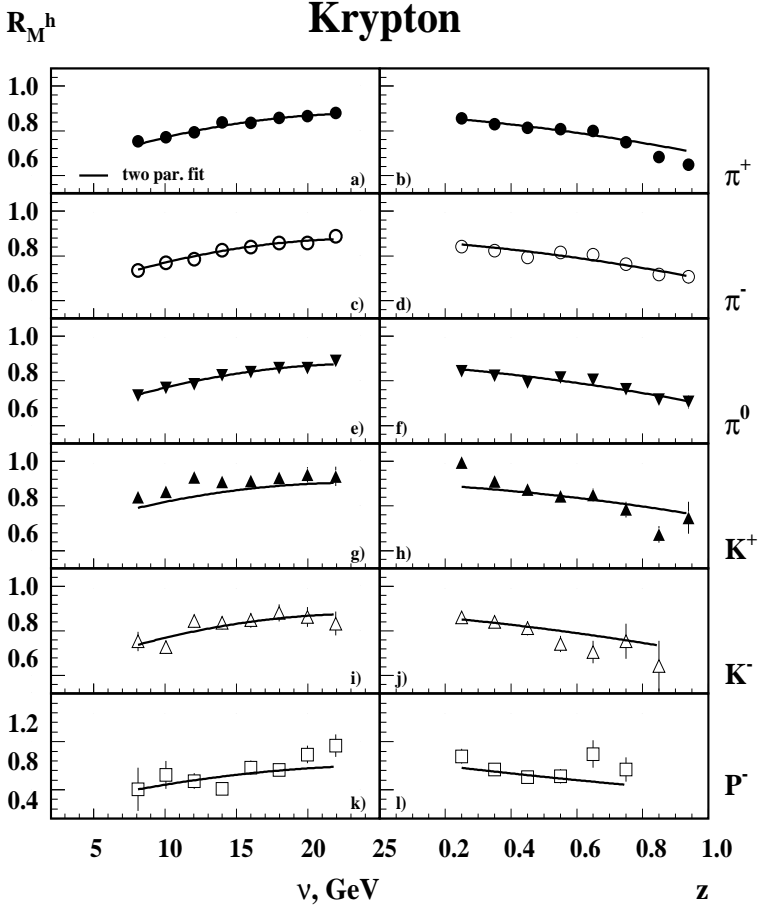}
\end{center}
\caption{\label{xx4}Hadron multiplicity ratio R of different species of hadrons produced on $^{84}$Kr
target~\cite{A15a} as a function of $\nu$ (left panel) and z (right panel). The curves are
calculated with the best fit 
parameters described in the caption of Fig. 3.}
\end{figure}
\begin{figure}
\begin{center}
\epsfxsize=10cm
\epsfysize=12cm
\epsfbox{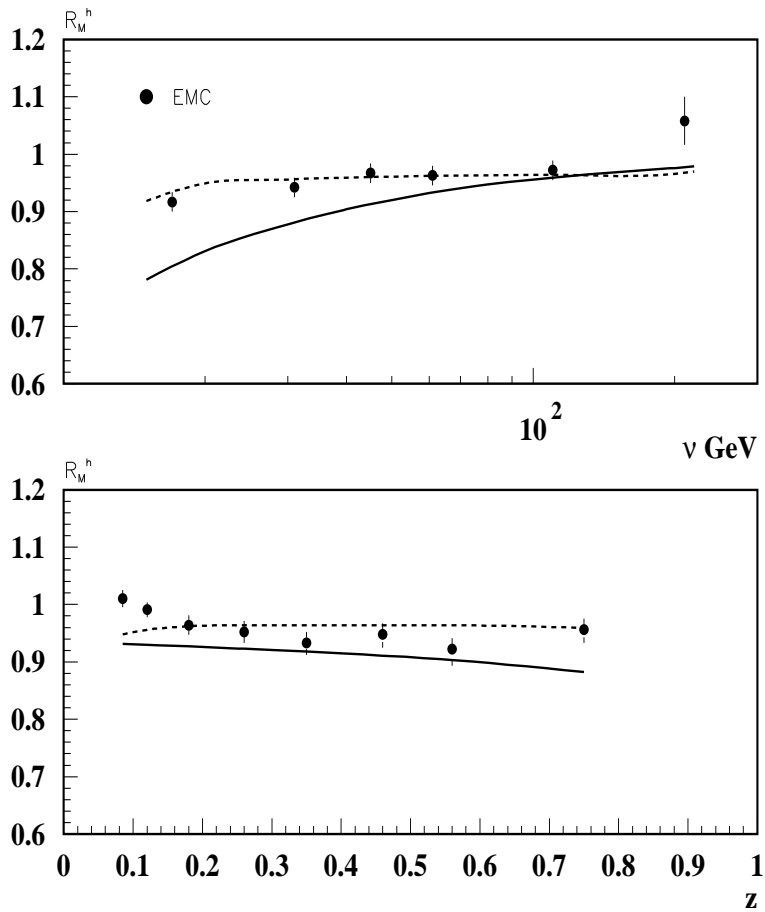}
\end{center}
\caption{\label{xx5}
Hadron multiplicity ratio R of charged hadrons for $^{63}$Cu nucleus as a function of $\nu$ (upper panel) and
z (lower panel). The dashed curves correspond to the Two-Scale Model, the solid ones to the improved version. 
The solid curves are calculated with the best set of parameters described in the caption of Fig. 3. The dashed
curves are calculated with $\tau_c$(4), NDF (19), $\sigma_q$ = 4.2~mb nad $\sigma_s$= 16.6~mb.}
\end{figure}
\begin{figure}
\begin{center}
\epsfxsize=10cm
\epsfysize=12cm
\epsfbox{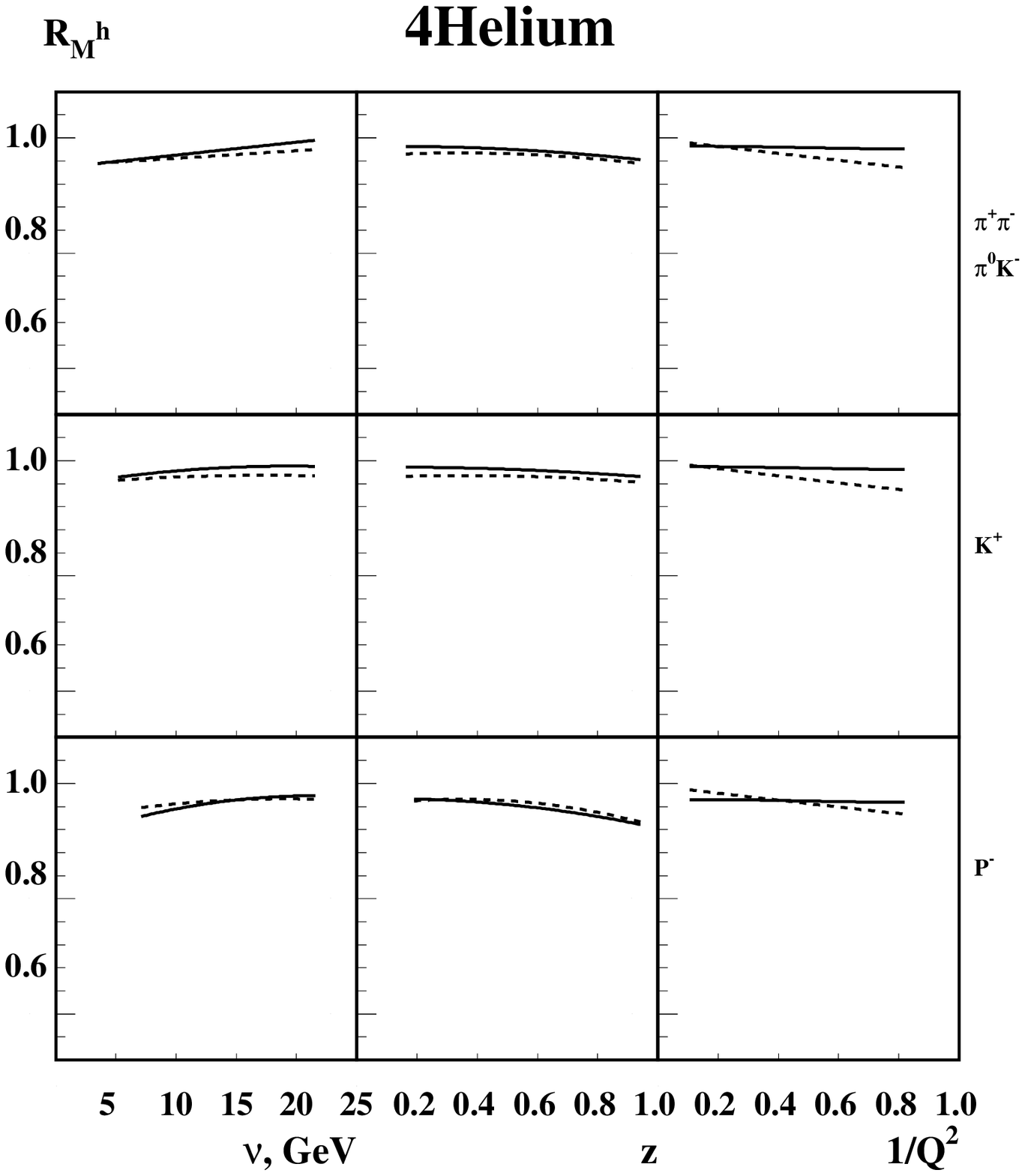}
\end{center}
\caption{\label{xx6}Hadron multiplicity ratio R of different species of hadrons produced on $^{4}$He
target as a function of $\nu$ (left panel), z (central panel) and Q$^2$ (right panel). The solid curves are the 
best fit using improved Two-Scale model with the parameters described in
the caption of Fig. 3. The dashed 
curves correspond to the 
best fit using the simple Two-Scale model
with $\tau_c$ defined in (4), NDF (17), $\sigma_q$ = 4.2~mb and $\sigma_s$ = 16.6~mb}
\end{figure}
\begin{figure}
\begin{center}
\epsfxsize=10cm
\epsfysize=12cm
\epsfbox{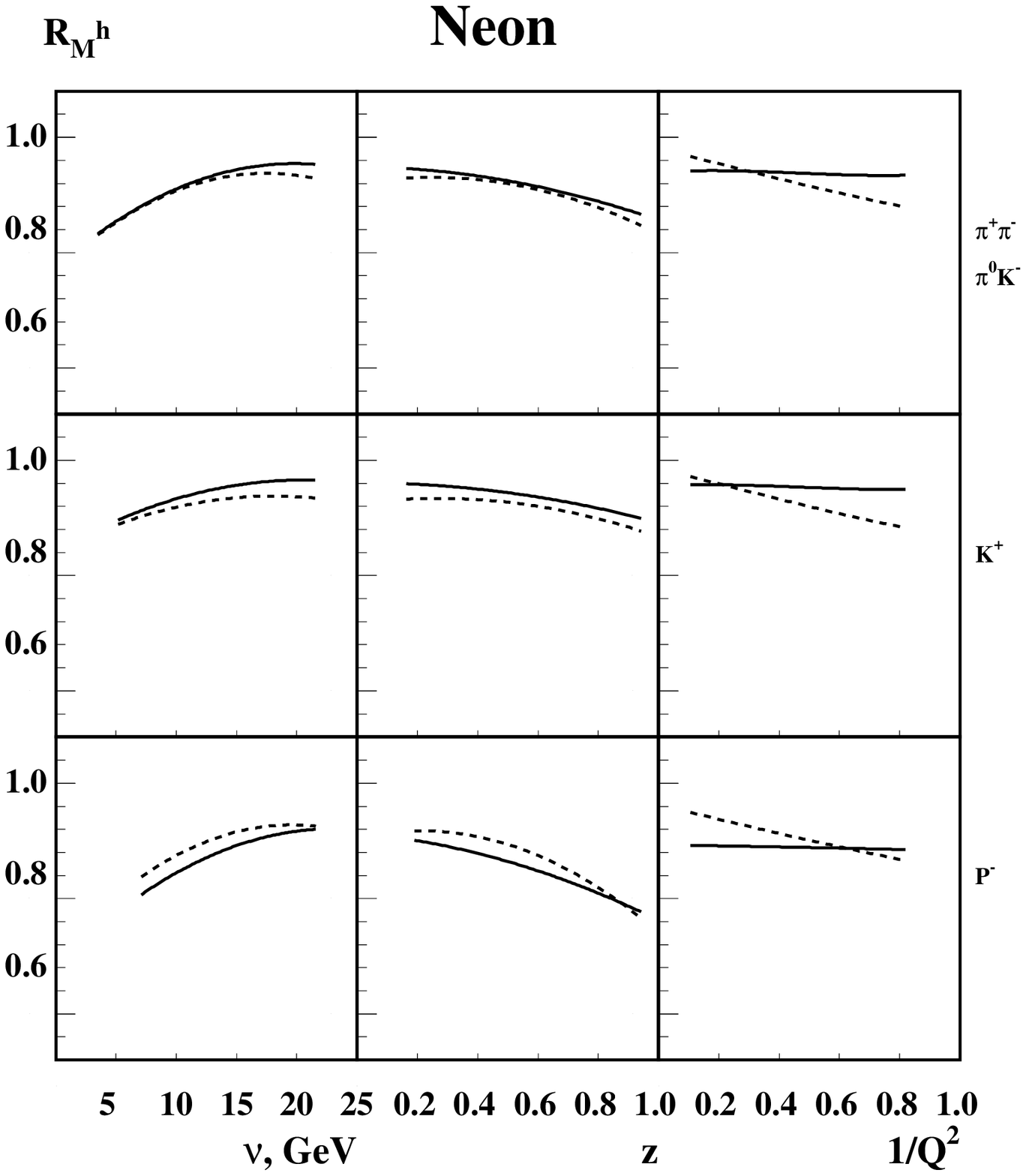}
\end{center}
\caption{\label{xx7}The same as described in the caption of the Fig. 6 done for $^{20}$Ne target.}
\end{figure}
\begin{figure}
\begin{center}
\epsfxsize=10cm
\epsfysize=12cm
\epsfbox{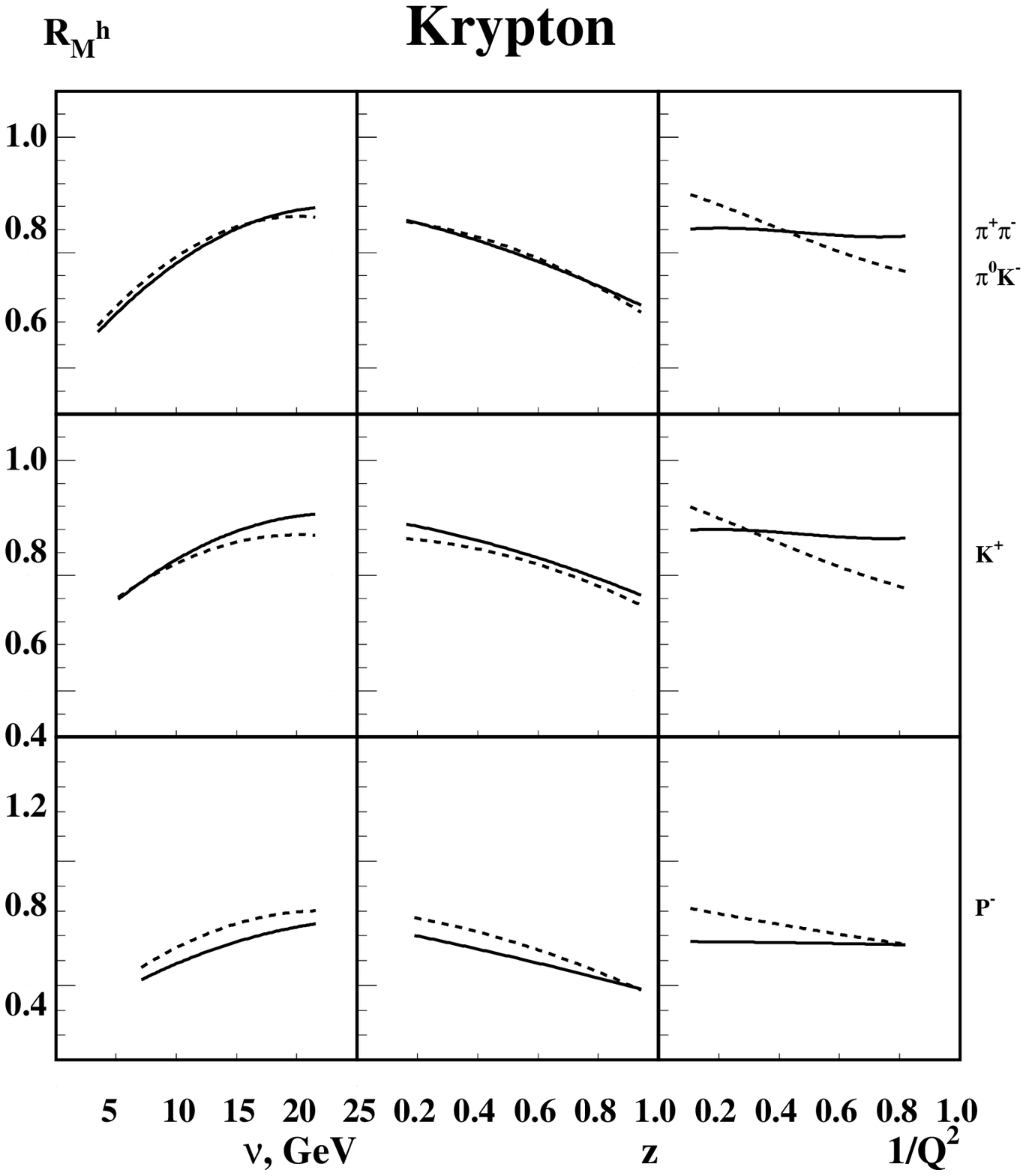}
\end{center}
\caption{\label{xx8} The same as described in the caption of the Fig. 6 done for $^{84}$Kr target.}
\end{figure}
 
 We represent the NA ratio as a function of inverse Q$^2$, because of connection
 of this dependence with the Higher Twist effects. Indeed, from the equations
 (9),(6), (12)-(15) and (1) we can conclude, that in first approximation the expansion over
 the degrees of 1/Q$^2$ for NA ratio can be represented in form $R_A=a+b/Q^2$, where b 
 is negative.
 
 
  One has to note, that for calculation of 1/Q$^2$-dependence, the
 $\sigma_q$(Q$^2$) was used instead of $\sigma_q$. Corresponding expression is
 given by (9).
We also take into account nuclear effects in deuterium. This
 means, that instead of a simple formula (1), we use for calculations 
 the ratio of (1) for nucleus to the (1) for deuterium.
 For deuterium as NDF we use Hard Core Deuteron Wave Functions
 from Ref.~\cite{A29}.
 
Using the best set of parameters obtained by fitting the published HERMES data~\cite{A15,A15a} we calculated
the predictions for the new set of the most precise in the world HERMES data~\cite{dis03} for 
$^4$He (Fig. 6),$^{20}$Ne (Fig. 7) and $^{84}$Kr (Fig. 8).

In order to demonstrate the achieved advantages for Imroved Two-Scale model not
only on the level of obtained $\chi^2$ values, one
can compare how these two versions are describing the NA data for pions on two
nuclear targets for z (see right panel of Fig. 9) and $\nu$ (see left panel of Fig. 9) dependencies. It's
clearly seen from this plot that being about the same for $\nu$ dependence these
two versions remarkable differ for z dependence.

\begin{figure}
\begin{center}
\epsfxsize=10cm
\epsfysize=12cm
\epsfbox{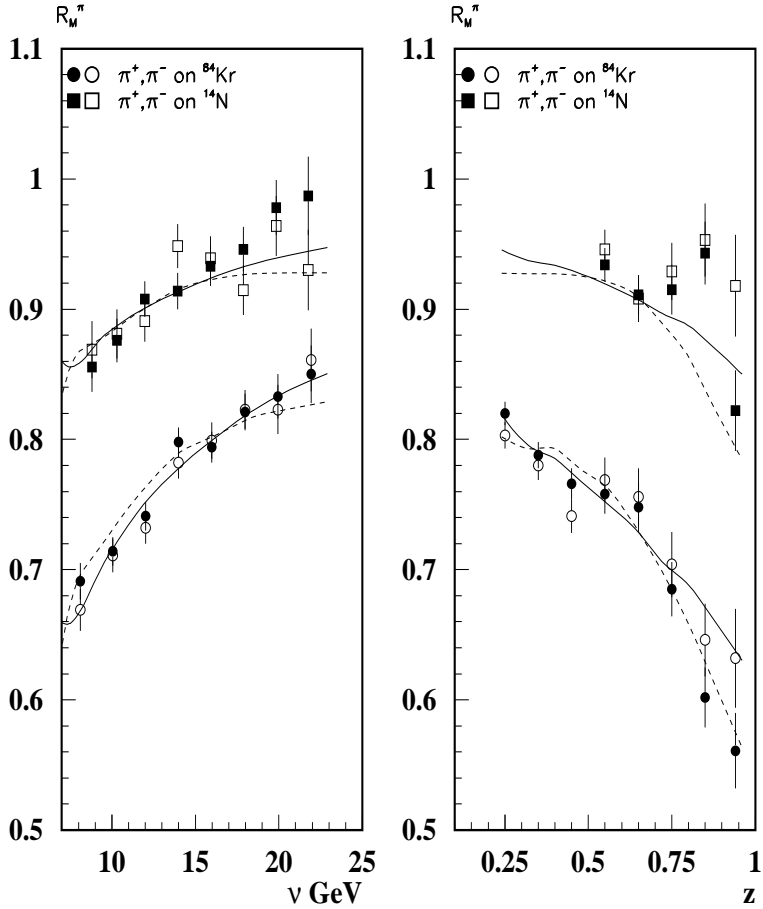}
\end{center}
\caption{\label{xx9} Descriptive ability of the Two-Scale model and its 
Improved version: right panel - for z, and left panel - for $\nu$ dependencies of NA. 
The solid lines on both panels correspond to the Improved
version, the dashed ones are for simple Two-Scale model.
}
\end{figure}
The last Figure 10 is related to the predictions, done for already presented by HERMES~\cite{dis03} data on 
$^4$He, $^{20}$Ne and $^{84}$Kr targets with the extended
kinematics, as well as for $^{131}Xe$ target, on
which the data is awaiting soon from the HERMES Collaboration. Two set of the
best fit parameters were fixed: one marked as a dashed curves on Fig. 10 is related
to the simple Two-Scale Model,  
next one, marked as a solid curves is related to the
Improved version of the Two-Scale Model. Left panel corresponds to z dependence
of NA for pions, right panel is related to the $\nu$ dependence of NA for pions.
We can note that again as for other nuclear targets, the difference in simple and improved versions is
remarkable for Xe in z dependence.
\begin{figure}
\begin{center}
\epsfxsize=10cm
\epsfysize=12cm
\epsfbox{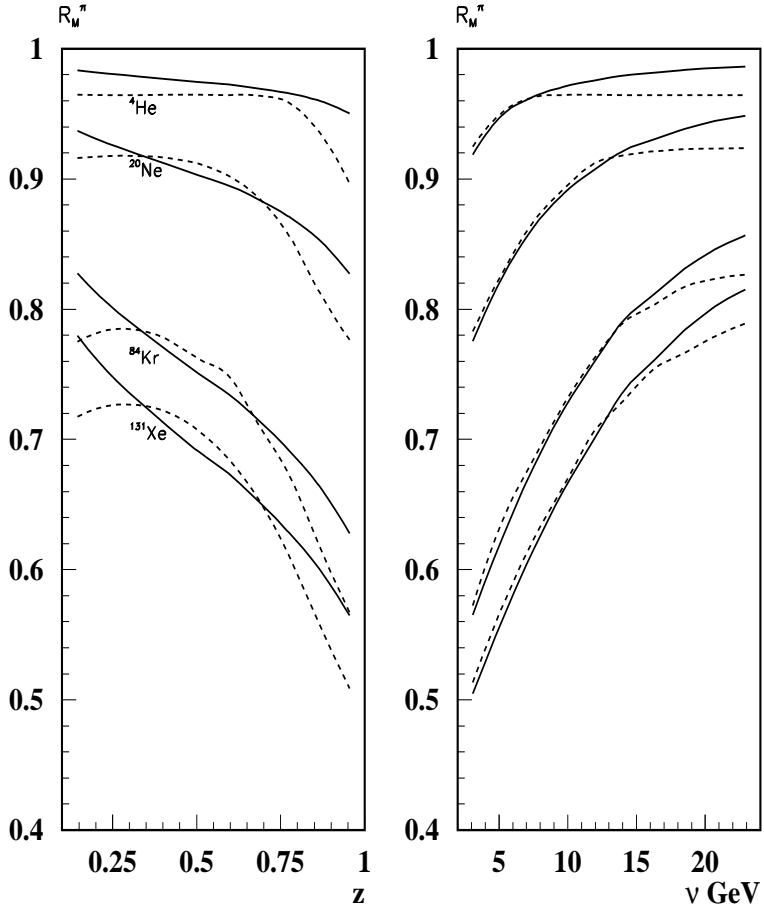}
\end{center}
\caption{\label{xx10} Two-Scale model (dashed lines) and its improved version 
(solid lines). The predictions for data on $^4$He, $^{20}$Ne, $^{84}$Kr and $^{131}$Xe
done for z and $\nu$ dependencies of NA.
}
\end{figure}
\begin{itemize}
\section{Conclusions.}
\item The HERMES data for $\nu$- and z - dependencies of nuclear attenuation 
of $\pi^+$ and $\pi^-$ mesons on two nuclear targets ($^{14}$N and $^{84}$Kr) were used to perform the fit of the Two-Scale Model and its Improved Version.
\item Criterion $\chi^2$ was used for the first time to analyse the nuclear attenuation data fit.
\item Two-parameter fit demonstrates satisfactory agreement to the HERMES data. Minimum $\chi^2$ (best fit) was obtained for 
improved Two-Scale Model, including expressions (12) for $\sigma^{str}$ and (3) for $\tau_c$.
The published HERMES data do not give the possibility to make a choice between expressions (12)-(15), as well as to prefere 
definition (3) or (4) for $\tau_c$, because they give close values of $\chi^2$. Preferable NDF's are set number one and two.
\item More precise data expected from HERMES~\cite{dis03} will provide essentially definite situation with the choice of preferable NDF, expressions for $\sigma^{str}$ and $\tau_c$.
\item In all versions we have obtained $\sigma_q \ll \sigma_h$. This indicates that at early stage of hadronization process 
Color Transparency takes place.
\end{itemize}
\vskip 0.5cm
\begin{tabular}{|c|c|c|c|c|c|c|} \hline
\multicolumn{7}{|c|}{\hskip 0.9cm $\tau_c$(3)\hskip 2.6cm \vline\vline \hskip 2.2cm $\tau_c$(4)}\\
\hline
NDF&$\sigma_q$ (mb)&$\sigma_s$ (mb)& $\chi^2$ /d.o.f.&$\sigma_q$ (mb)&$\sigma_s$ (mb)& $\chi^2$ /d.o.f.\\
\hline\hline
1&5.3$\pm$0.01&17.1$\pm$0.08&4.3&4.2$\pm$0.01&16.6$\pm$0.07&2.3 \\
\hline
2&5.5$\pm$0.01&17.7$\pm$0.08&4.5&4.3$\pm$0.01&17.3$\pm$0.07&2.4 \\ 
\hline
3&5.8$\pm$0.01&18.3$\pm$0.08&4.8&4.4$\pm$0.01&18.1$\pm$0.07&2.6 \\
\hline
\end{tabular}
\vskip 0.5cm
Table 1. The Two-Scale Model. Best values for fitted parameters and 
$\chi^2$/d.o.f.\\ (N$_{exp}$=58, N$_{par}$=2)

\vskip 0.5cm

\begin{tabular}{|c|c|c|c|c|c|c|} \hline
\multicolumn{7}{|c|}{\hskip 0.9cm $\sigma_{str}$(12)\hskip 2.6cm \vline\vline \hskip 2.2cm$\sigma_{str}$(13)}\\
\hline
NDF&$\sigma_q$ (mb)&$c$& $\chi^2$ /d.o.f.&$\sigma_q$ (mb)&$c$& $\chi^2$ /d.o.f.\\
\hline\hline
1&0.46$\pm$0.02&0.32$\pm$0.03&1.4&3.5$\pm$0.01&0.23$\pm$0.002&1.9 \\
\hline
2&0.62$\pm$0.01&0.31$\pm$0.01&1.7&3.7$\pm$0.01&0.22$\pm$0.02&2.1 \\ 
\hline
3&0.78$\pm$0.02&0.30$\pm$0.03&1.8&3.9$\pm$0.01&0.21$\pm$0.003&2.3 \\
\hline
\end{tabular}
\vskip 0.3cm

\begin{tabular}{|c|c|c|c|c|c|c|} \hline
\multicolumn{7}{|c|}{\hskip 0.9cm $\sigma_{str}$(14)\hskip 2.6cm \vline\vline \hskip 2.2cm$\sigma_{str}$(15)}\\
\hline
NDF&$\sigma_q$ (mb)&$c$& $\chi^2$ /d.o.f.&$\sigma_q$ (mb)&$c$& $\chi^2$ /d.o.f.\\
\hline\hline
1&1.1$\pm$0.01&0.15$\pm$0.03&2.1&3.7$\pm$0.01&0.15$\pm$0.02&2.3 \\
\hline
2&1.3$\pm$0.02&0.15$\pm$0.03&2.4&3.9$\pm$0.01&0.14$\pm$0.02&2.6 \\ 
\hline
3&1.5$\pm$0.02&0.14$\pm$0.03&2.8&4.1$\pm$0.01&0.14$\pm$0.02&2.9 \\
\hline
\end{tabular}
\vskip 0.5cm
Table 2a. The Improved Two-Scale Model:$\tau_c$(3). 
Best values for fitted parameters and 
$\chi^2$/d.o.f. (N$_{exp}$=58, N$_{par}$=2). 
\vskip 1cm
\newpage
\begin{tabular}{|c|c|c|c|c|c|c|} \hline
\multicolumn{7}{|c|}{\hskip 0.9cm $\sigma_{str}$(12)\hskip 2.6cm \vline\vline \hskip 2.2cm$\sigma_{str}$(13)}\\
\hline
NDF&$\sigma_q$ (mb)&$c$& $\chi^2$ /d.o.f.&$\sigma_q$ (mb)&$c$& $\chi^2$ /d.o.f.\\
\hline\hline
1&0.0$\pm$0.001&0.56$\pm$0.02&4.6&0.97$\pm$0.01&0.17$\pm$0.002&1.6 \\
\hline
2&0.0$\pm$0.002&0.53$\pm$0.02&4.3&1.0$\pm$0.02&0.17$\pm$0.02&1.5 \\ 
\hline
3&0.0$\pm$0.002&0.49$\pm$0.006&4.0&1.1$\pm$0.02&0.16$\pm$0.02&1.6 \\
\hline
\end{tabular}
\vskip 0.3cm
\begin{tabular}{|c|c|c|c|c|c|c|} \hline
\multicolumn{7}{|c|}{\hskip 0.9cm $\sigma_{str}$(14)\hskip 2.6cm \vline\vline \hskip 2.2cm$\sigma_{str}$(15)}\\
\hline
NDF&$\sigma_q$ (mb)&$c$& $\chi^2$ /d.o.f.&$\sigma_q$ (mb)&$c$& $\chi^2$ /d.o.f.\\
\hline\hline
1&0.0$\pm$0.001&0.24$\pm$0.02&3.0&1.5$\pm$0.02&0.103$\pm$0.02&1.5 \\
\hline
2&0.0$\pm$0.002&0.21$\pm$0.02&2.9&1.7$\pm$0.02&0.096$\pm$0.02&1.6 \\ 
\hline
3&0.0$\pm$0.002&0.18$\pm$0.02&2.8&1.8$\pm$0.02&0.089$\pm$0.02&1.8 \\
\hline
\end{tabular}
\vskip 0.5cm
Table 2b. The Improved Two-Scale Model:$\tau_c$(4). Best values for fitted 
parameters and 
$\chi^2$/d.o.f. (N$_{exp}$=58, N$_{par}$=2).
\vskip 1cm
We do not include in consideration NA of protons, because in this
 case additional mechanisms connected with color interaction (string-
 flip) and final hadron rescattering become essential (see for instance
 Ref.~\cite{A3, A5})
 
We would like to acknowledge P. Di Nezza and E. Aschenauer for helpful discussions and suggestions. We also thank G.
Elbakian as well as many other colleagues from the HERMES Collaboration for fruitful discussions. 


\begin{thebibliography}{99}
\bibitem{A1} N.Nikolaev, Z.Phys. {\bf C5}(1980)291; V.Anisovich et al., 
Nucl.Phys.
 {\bf B133} (1978) 477; G.Davidenko and N.Nikolaev, Nucl.Phys. {\bf B135} (1978) 333
\bibitem{A2} A.Bialas, Acta Phys.Pol. {\bf B11} (1980) 475
\bibitem{A3} M.Gyulassy and M.Plumer, Nucl.Phys. {\bf B346} (1990) 1
\bibitem{A4} J.Ashman et al., Z.Phys. {\bf C52}( 1991) 1
\bibitem{A5} J.Czyzewski and P.Sawicki, Z.Phys. {\bf C56} (1992) 493
\bibitem{A6} R.Badalyan, Z.Phys. {\bf C55} (1992) 647
\bibitem{A7} N.Akopov, G.Elbakian, L.Grigoryan, hep-ph/0205123(2002)
\bibitem{A8}  T. Falter, W. Cassing, K. Gallmeister, and U. Mosel,
Phys. Lett. B 594, 61 (2004);\\ 
T. Falter, W. Cassing, K. Gallmeister, and U. Mosel, Phys. Rev. C (in press),\\ 
nucl-th/0406023 
\bibitem{A9} J.Dias De Deus, Phys.Lett. {\bf B166}(1986) 98
\bibitem{A10} A.Accardi, V.Muccifora, H.J.Pirner, Nucl.Phys.,{\bf A720} (2003) 131;\\
 nucl-th/0211011(2002)
\bibitem{A11} F.Arleo, JHEP {\bf 11}(2002)44; hep-ph/0210105 (2002)
\bibitem{A12} X.-N.Wang and X.Guo, Nucl.Phys. {\bf A696} (2001) 788;
 E.Wang and X.-N.Wang, Phys.Rev.Lett. {\bf 89}(2002) 162301
\bibitem{A13} B.Kopeliovich, J.Nemchik and E.Predazzi, Proceedings of the
 workshop on Future Physics at HERA, Edited by G.Ingelman, A. De
 Roeck and R.Klanner, DESY, 1995/1996,{\bf vol.2}, p.1038 (nucl-th/9607036);
 B.Kopeliovich et al., hep-ph/0311220 (2003)
\bibitem{A14} D.J.Dean et al., Phys.Rev. {\bf C46} (1992) 2066
\bibitem{A15} A.Airapetian et al., Eur.Phys.J. {\bf C20} (2001) 479;
\bibitem{A15a} A.Airapetian et al.,  Phys. Lett. {\bf B577} (2003) 37-46;
\bibitem{dis03}  G.Elbakyan [HERMES Collaboration], in Proceedings of DIS 2003,
St.Petersburg, 2003, edited by V.T.Kim and L.N.Lipatov, p.597
\bibitem{A16} B.Kopeliovich, Phys.Lett. {\bf B243} (1990) 141
\bibitem{A17} A.Bialas, M.Gyulassy, Nucl.Phys. {\bf B291} (1987) 793;
 T.Chmaj, Acta Phys.Pol. {\bf B18} (1987) 1131
\bibitem{A18} B.Andersson et al., Phys.Rep. {\bf 97} (1983) 31
\bibitem{A19} B.Kopeliovich, B.Povh, Proceedings of the
 workshop on Future Physics at HERA, Edited by G.Ingelman, A. De
 Roeck and R.Klanner, DESY, 1995/1996,{\bf vol.2}, p.959
\bibitem{A20} K.Golec-Biernat et al., Phys.Rev. {\bf D59} (1998) 014017
\bibitem{A21} B.Kopeliovich et al., "Hadron Structure '92" Proceedings
 Stara Lesna, Czecho-Slovakia, Sept.6-11, 1992, p.164
\bibitem{A22} B.Kopeliovich, J.Nemchik, preprint JINR E2-91-150 (1991);
 preprint of INFN-ISS 91/3(1991) Roma
\bibitem{A23} G.Farrar et al.,  Phys.Rev.Lett. {\bf 61} (1988) 686
\bibitem{A24} T.Sjostrand, L.Lonnblad, S.Mrenna, hep-ph/0108264 (2001);
 LU TP 01-21
\bibitem{A25} L.Elton, "Nuclear Sizes" Oxford University Press, 1961, p.34
\bibitem{A26} A.Bialas et al., Phys.Lett. {\bf B133} (1983) 241
\bibitem{A27} A.Bialas et al., Nucl.Phys. {\bf B291} (1987) 793
\bibitem{A28} A.Capella et al., Phys.Rev. {\bf D18} (1977) 3357
\bibitem{A29} R.V.Reid, Annals of Physics {\bf 50} (1968) 411
\end{thebibliography}
\end{document}